\documentclass[a4paper,12pt]{article}
\usepackage{graphicx}
\usepackage[top=3cm, bottom=3cm, left=3cm, right=3cm]{geometry}
\usepackage{amssymb,amsmath,amsthm}
\usepackage{mathrsfs} %script fonts
\usepackage{hyperref}
\usepackage[normalem]{ulem} % for \sout{...}
\usepackage[applemac]{inputenc} % for dealing with oddities of the TeX-shop text-editor...
\title{\textbf{Surface gravities for non-Killing horizons}}
\author{Bethan Cropp$^{1,2}$, Stefano Liberati$^{1,2}$, and Matt Visser$^3$
\\[10pt] 
$^1$ SISSA, International School for Advanced Studies, \\
Via Bonomea 265, I-34136 Trieste, Italy
\\[10pt]
$^2$ INFN, Sezione di Trieste, Trieste, Italy
\\[10pt]
$^3$ School of Mathematics, Statistics, and Operations Research; \\
Victoria University of Wellington; PO Box 600; \\
Wellington 6140; New Zealand.
}
\date{11 February 2013; \LaTeX-ed \today}
\begin{document}
%------------------------------------------------------------------------------------------------------------------------------------------
% very standard definitions
%------------------------------------------------------------------------------------------------------------------------------------------
\def\d{{\mathrm{d}}}
\newcommand{\scri}{\mathscr{I}}
\newcommand{\sun}{\ensuremath{\odot}}
\def\J{{\mathscr{J}}}
\def\sech{{\mathrm{sech}}}
\def\T{{\mathcal{T}}}
\def\tr{{\mathrm{tr}}}
\def\diag{{\mathrm{diag}}}
\def\ln{{\mathrm{ln}}}
% very standard definitions
%------------------------------------------------------------------------------------------------------------------------------------------
\def\v{\mathbf{v}}
\def\n{\mathbf{n}}
\def\x{\mathbf{x}}
\def\s{\mathbf{s}}
%\def\0{\mathbf{0}}
%------------------------------------------------------------------------------------------------------------------------------------------
\maketitle
%------------------------------------------------------------------------------------------------------------------------------------------
\begin{abstract}
%------------------------------------------------------------------------------------------------------------------------------------------
\noindent
There are many logically and computationally distinct characterizations of the surface gravity of a horizon, just as there are many logically rather distinct notions of horizon. 
Fortunately, in standard general relativity, for stationary horizons, most of these characterizations are degenerate.  
However, in modified gravity, or in analogue spacetimes, horizons may be non-Killing or even non-null, and hence these degeneracies can be lifted.
We present a brief overview of the key issues, 
specifically focusing on horizons in analogue spacetimes and universal horizons in modified gravity.

%------------------------------------------------------------------------------------------------------------------------------------------
 Killing horizons. non-Killing horizons. rigidity theorems. modified gravity. analogue spacetimes. Hawking radiation. 
%------------------------------------------------------------------------------------------------------------------------------------------

\bigskip
\noindent
E-mail: {\sf liberati@sissa.it}, {\sf  bcropp@sissa.it}, {\sf matt.visser@msor.vuw.ac.nz}

%------------------------------------------------------------------------------------------------------------------------------------------
\end{abstract}
%------------------------------------------------------------------------------------------------------------------------------------------
\clearpage
\bigskip
\hrule
\tableofcontents
\bigskip
\hrule
%---------------------------------------------------------------------------------------------------------------------
\def\L{\mathcal{L}}
%------------------------------------------------------------------------------------------------------------------------------------------
\clearpage
%------------------------------------------------------------------------------------------------------------------------------------------
\section{Introduction}
%------------------------------------------------------------------------------------------------------------------------------------------

Surface gravity is an important quantity in classical general relativity, which plays a vital role in black hole thermodynamics and semi-classical aspects of gravity, being closely related to the temperature of Hawking radiation. 
However,  in a large number of situations, the surface gravity cannot be calculated unambiguously, as standard definitions rely on the existence of a stationary spacetime with a Killing horizon.

In recent years, a quite significant amount of work has been devoted to considering extensions to the usual notion of surface gravity that would be suitable for dynamical situations in standard general relativity, such as a forming or evaporating black hole (see, for instance~\cite{Nielsen:2005af, Nielsen:2007ac, Pielahn:2011ra} and~\cite{Hayward:1993wb, Fodor:1996rf, Hayward:1997jp,  Booth:2003ji, Booth:2006bn}). 
Much less effort has been devoted to stationary scenarios where the horizon is no longer a Killing horizon. 
The explanation for this is simple: 
For the standard case of general relativity, due to the rigidity theorem (see, for instance~\cite{HawkingEllis, carter1, carter2} and~\cite{Heusler:1998, Friedrich:1998wq, Robinson:2004, Kodama:2011}), in stationary spacetimes all event horizons are automatically Killing horizons (\emph{i.e.}, the spacetime must possess a Killing field which is normal to the event horizon). 
However, this result hinges on the Einstein field equations, and in modified gravity, or in the arena of analogue spacetimes, there is no \emph{a priori} reason to expect this result will continue to hold.
We will address a number of scenarios where the standard calculations for surface gravities either will not hold, or will give rise to distinct quantities. 

This technical heart of the paper is essentially divided into three sections. In the first section, we will briefly present the standard general relativity case, initially making a foray into dynamical situations to demonstrate how the different definitions of surface gravity can diverge, (though they will asymptotically agree in the adiabatic limit), and subsequently run through several quite standard ways to calculate the surface gravity in stationary spacetimes, as presented (for instance) by Wald in reference~\cite{Wald}, drawing explicit attention to the assumptions built into the calculations; assumptions that we shall then relax in subsequent discussion. 
As a first step in this relaxation process we consider the conformal Killing horizons of Jacobson and Kang~\cite{Jacobson-Kang}. 

The second section is devoted to the analogue spacetime case, focussing specifically on acoustic horizons. 
In this context, all horizons are null surfaces, (in fact, they are even geodesic null surfaces), but in the case of non-zero rotation, (non-zero vorticity, or more precisely non-zero helicity), can nevertheless be non-Killing. 
We demonstrate that the different definitions of the surface gravity will in this context lead to physically and mathematically distinct quantities, and discuss which is the most relevant one in the case of analogue horizon thermodynamics.

The third section will be devoted to discussing a new class of horizons, the so-called ``universal horizons'',  recently discovered in theories with Lorentz violation, such as Einstein-aether and Horava--Lifshitz gravity.  (See~\cite{Berglund:2012fk, Berglund:2012bu, Barausse:2011pu, Barausse:2012ny, Barausse:erratum, Blas:2011ni}.) Such horizons are spacelike instead of null surfaces, and are not Killing horizons. The physics is quite different from what one might otherwise expect.

Finally we end with a brief discussion putting our calculations in context. 
In particular, while for definiteness in this paper we will discuss non-Killing horizons in analogue spacetime and  Einstein-aether and Horava--Lifshitz contexts, the issues raised are much more general --- similar considerations will apply in various modified gravity models where modification of the Einstein equations generically eliminates the rigidity theorems so non-Killing horizons are likely to be generic. 
For instance, non-Killing horizons have very recently become of interest both in AdS/CFT~\cite{Fischetti:2012vt} and holographic~\cite{Figueras:2012rb} situations.

%------------------------------------------------------------------------------------------------------------------------------------------
\begin{table}[!htdp]
%------------------------------------------------------------------------------------------------------------------------------------------
\caption{Some of the multiple notions of surface gravity}. 
\vspace{-15pt}
\begin{center}
\begin{tabular}{|c| |c|}
\hline
\hline
Name & Key features \\
\hline
\hline
peeling & peeling off properties for null geodesics near horizon\\
\hline
inaffinity & inaffinity properties for null geodesics on horizon\\
\hline
\hline
normal & null normal to a null surface\\
\hline
generator & anti-symmetrized derivatives of horizon generators\\
\hline
tension & tension in an ideal massless rope\\
\hline
expansion & geodesic expansion transverse to the horizon\\
\hline
\hline
Euclidean & elimination of angle deficit at horizon\\
\hline
\hline

\end{tabular}
\\[5pt]
Some of these definitions require specific simplifying assumptions. \\
Others are (or can be made to be) more general. \\
All definitions are equivalent for Killing horizons. 
\end{center}
\label{default}
%------------------------------------------------------------------------------------------------------------------------------------------
\end{table}%
%------------------------------------------------------------------------------------------------------------------------------------------

%------------------------------------------------------------------------------------------------------------------------------------------
\section[Standard general relativity --- peeling off versus inaffinity]{Standard general relativity:\\
 Peeling off versus inaffinity}
%------------------------------------------------------------------------------------------------------------------------------------------

Even in standard general relativity, which is one of the simplest frameworks one might envisage, there are essentially two basic conceptions of surface gravity, related to the inaffinity of null geodesics \emph{on the horizon}, and the the peeling off properties of null geodesics \emph{near the horizon}, respectively. 
For stationary Killing horizons these two notions coincide, but even in the simplest case of a spherically symmetric dynamical evolution these are two quite distinct quantities. We will work thorough a brief calculation, adapted from~\cite{Barcelo:2010xk} (see also~\cite{Barcelo:2010pj}), as an example.
Without loss of generality, write the metric in the form
\begin{equation}
\d s^2 = - e^{-2\Phi(r,t)} [1-2m(r,t)/r)] \d t^2 + {\d r^2\over1-2m(r,t)/r} + r^2 \{ \d\theta^2+\sin^2\theta\; \d\phi^2\} ,
\end{equation}
and define the ``evolving horizon'', $r_H(t)$,  by the location where $2m(r,t)/r = 1$. 
(Working from the Kodama vector, a ``geometrically natural'' justification for interest in this particular form of the line element is presented in~\cite{Abreu:2010ru}.)

%------------------------------------------------------------------------------------------------------------------------------------------
\subsection{Peeling off properties of null geodesics}\label{peeling}
%------------------------------------------------------------------------------------------------------------------------------------------

A radial null geodesic satisfies
\begin{equation}
\left({\d r\over\d t}\right) =  \pm e^{-\Phi(r,t)} [1-2m(r,t)/r)].
\end{equation}
If the geodesic is near $r_H(t)$, that is  $r \approx r_H(t)$, then we can Taylor expand
 \begin{equation}
{\d r\over\d t} =  \pm {e^{-\Phi(r_H(t),t)} [1-2m'(r_H(t),t)]\over r_H(t)} \; [r(t) - r_H(t)] +{\cal O}\left( [r(t) - r_H(t)]^2 \right),
\end{equation}
where the dash indicates a radial derivative. That is, \emph{defining}
\begin{equation}
\kappa_\mathrm{peeling}(t) =  {e^{-\Phi(r_H(t),t)} [1-2m'(r_H(t),t)]\over 2 r_H(t)},
\end{equation}
which, in the static case, reduces to the standard result~\cite{Visser:1992qh}
\begin{equation}
\kappa = {e^{-\Phi_H} (1-2m'_H)\over 2 r_H},
\end{equation}
we have
 \begin{equation}
{\d r\over\d t} =  \pm 2 \kappa_\mathrm{peeling}(t) \; [r(t) - r_H(t)] +{\cal O}\left( [r(t) - r_H(t)]^2 \right).
\end{equation}
Then, for two null geodesics $r_1(t)$ and $r_2(t)$ on the \emph{same} side of the evolving horizon
\begin{equation}
{\d |r_1-r_2|\over\d t} \approx 2 \kappa_\mathrm{peeling}(t) \; |r_1(t) - r_2(t)|,
\end{equation}
(automatically keeping track of all the signs), so
\begin{equation}
|r_1(t)-r_2(t)| \approx |r_1(t_0)-r_2(t_0)| \; \exp \left[ 2\int \kappa_\mathrm{peeling}(t) \d t \right].
\end{equation}
This makes manifest the fact that $\kappa_\mathrm{peeling}$ as we have defined it is related to the exponential peeling off properties of null geodesics \emph{near the horizon}.

%------------------------------------------------------------------------------------------------------------------------------------------
\subsection{Inaffinity properties of null geodesics}
%------------------------------------------------------------------------------------------------------------------------------------------

Consider the outward-pointing radial null vector field
\begin{equation}
\ell^a = \left( 1,  e^{-\Phi(r,t)} (1-2m(r,t)/r), 0, 0\right).
\end{equation}
In a static spacetime, this null vector field is very simply related to the Killing vector, 
\begin{equation}
\ell^a = \chi^a + \epsilon^a{}_b \chi^b,
\end{equation}
where $\epsilon_{ab}$ is a 2-form acting on the $r$--$t$ plane, normalized by $\epsilon^{ab} \, \epsilon_{ab} = - 2$. The radial null vector field $\ell^a$ is automatically geodesic. 
Hence the inaffinity  $\kappa_\mathrm{inaffinity}(r,t)$ can be defined by 
\begin{equation}\label{radialinaffinity}
\ell^a \nabla_a \ell^b = 2 \kappa_\mathrm{inaffinity}(r,t) \; \ell^b,
\end{equation}
which \emph{always} exists, everywhere throughout the spacetime. 
This construction naturally extends the notion of on-horizon geodesic inaffinity, defined in a static spacetime as
\begin{equation}
\chi^a \nabla_a \chi^b = \kappa_\mathrm{inaffinity} \; \chi^b.
\end{equation}
That is,  equation (\ref{radialinaffinity}) naturally defines a notion of surface gravity even for a time-dependent geometry.
A brief calculation shows that at the evolving horizon~\cite{Nielsen:2005af, Abreu:2010ru},
\begin{eqnarray}
\kappa_\mathrm{inaffinity}(r_H(t),t) &=& {e^{-\Phi(r_H(t),t)} [1 - 2 m'(r_H(t),t)] \over 2r}  - {1\over2}\dot\Phi(r_H(t),t),  \nonumber\\
&=& \kappa_\mathrm{peeling}(t)  - {1\over2}\dot\Phi(r_H(t),t) .
\end{eqnarray}
While we do not \emph{a priori} know exactly where the event horizon (absolute horizon) is, we can certainly assert that when asymptotically approaching a  quasi-static situation the event horizon will be close to the evolving horizon. We then have
\begin{equation}
r_E(t) \approx r_H(t),
\end{equation}
in which case we can expand in a Taylor series
\begin{equation}
\kappa_\mathrm{inaffinity}(r_E(t),t)\approx  \kappa_\mathrm{inaffinity}(r_H(t),t) +  \kappa_\mathrm{inaffinity}'(r_H(t),t) [r_E(t)-r_H(t)].
\end{equation}
That is
\begin{equation}
\kappa_\mathrm{inaffinity}(r_E(t), t)\approx   \kappa_\mathrm{peeling}(t)  - {1\over2}\dot\Phi(r_H(t),t)   +  \kappa_\mathrm{inaffinity}'(r_H(t),t) [r_E(t)-r_H(t)].
\end{equation}
In particular, for sufficiently slowly evolving horizons the two concepts are for all practical purposes indistinguishable. 

%------------------------------------------------------------------------------------------------------------------------------------------
\paragraph{Summary:}
%------------------------------------------------------------------------------------------------------------------------------------------

In general (even for spherical symmetry in standard general relativity)  $\kappa_\mathrm{peeling}(t)\neq  \kappa_\mathrm{inaffinity}(r,t)$, with strict equality only on the horizon, and only in the static case. 
This distinction is important, because it seems to be the peeling notion that is more closely associated with Hawking radiation~\cite{Barcelo:2010xk, Barcelo:2010pj}. 

%------------------------------------------------------------------------------------------------------------------------------------------
\section{Standard general relativity --- stationary case}\label{standardcase}
%------------------------------------------------------------------------------------------------------------------------------------------
Let us now consider stationary horizons in standard general relativity, so that (in view of the classical rigidity theorems) all horizons are automatically Killing. 
\begin{itemize}

\item 
The peeling definition of surface gravity $\kappa_\mathrm{peeling}$ is somewhat messy to write down in the general stationary case, though it is already clear from the spherically symmetric discussion above that it will almost certainly equal $\kappa_\mathrm{inaffinity}$.

\item
In contrast, for stationary horizons the inaffinity definition of surface gravity is typically restricted to an explicitly on-horizon version, and given by a simple explicit formula. In terms of the Killing vector $\chi$ (see for example Wald~\cite{Wald}):
\begin{equation}
\chi^a \nabla_a \chi^b = \kappa_\mathrm{inaffinity} \; \chi^b,
\end{equation}
where this formula now makes sense only on the horizon.

\item
A third notion of surface gravity is that of the null normal derivative evaluated on the horizon (see for example Wald~\cite{Wald}):
\begin{equation}
\nabla^a(\chi^b\chi_b)=-2\kappa_\mathrm{normal}\;  \chi^a.
\end{equation}
Equivalently, 
\begin{equation}
\chi^b \nabla_a \chi_b = - \kappa_\mathrm{normal}\; \chi_a.
\end{equation}
Using Killing's equation we see $ \kappa_\mathrm{normal}= \kappa_\mathrm{inaffinity} $, but this equality will generically fail once we move to consider non-Killing horizons. 
(We shall exhibit explicit failure of this equality for acoustic horizons later on in the article.)

\item
As a fourth notion of surface gravity Wald~\cite{Wald} furthermore argues that it is useful to define the equivalent of
\begin{equation}
\kappa_\mathrm{generator}^2 = -{1\over2} (\nabla^{\left[ a\right. } \chi^{\left. b\right] }) (\nabla_{\left[ a\right. } \chi_{\left. b\right] }),
\end{equation}
(this name is chosen because the integral curves of the vector field $\chi^a$ generate the horizon.)
This definition makes sense everywhere throughout the spacetime. 
A brief calculation~\cite{Wald} demonstrates that on the (Killing) horizon
\begin{equation}
\left.\kappa_\mathrm{generator}\right|_H = \kappa_\mathrm{inaffinity} .
\end{equation}
Again,  this inequality will generically fail once we move to consider non-Killing horizons.
(Also in this case we shall exhibit explicit failure of this equality for acoustic horizons later on in the article.)

\item
A fifth notion of surface gravity can be formulated in terms of the tension in an ideal massless rope holding a unit mass steady just above the Killing horizon:
\begin{equation}
\kappa_\mathrm{tension} = \lim_H \sqrt{-\chi^2}\, \Vert A\Vert .
\end{equation}
Here $\Vert A\Vert $ denotes the magnitude of the 4-acceleration. 
Wald demonstrates that for Killing horizons $\kappa_\mathrm{tension} =  \kappa_\mathrm{generator} =\kappa_\mathrm{inaffinity}$, but this equality will again generically fail once we move to consider non-Killing horizons.
(Again, we shall demonstrate explicit failure of this equality for acoustic horizons later on in the article.)

\item
A sixth notion of surface gravity recently developed by Jacobson and Parentani  is based on relating the surface gravity to the expansion of the 2-d surface drawn by (timelike) geodesic congruences orthogonal to the horizon. Define
\begin{equation}
\label{E:theta-0}
\theta_{2d} = h^a{}_b \, \nabla_a u^b,
\end{equation}
for $h^a{}_b$ the surface projector onto the 2-d surface generated by the congruence. We pick an appropriate congruence such that
\begin{equation}
\chi^a \nabla_a u^b=u^a\nabla_a \chi^b,
\end{equation}
and hence we can write this 2-d expansion as 
\begin{equation}
\theta_{2d}  =\frac{\frac{1}{2} u^a\nabla_a\chi^2}{\chi^2-(\chi\cdot u)^2}.
\end{equation}
Then on-horizon, where $\chi^2=0$, we have
\begin{equation}
\left.\theta_{2d}\right|_H  = -\frac{\frac{1}{2} u^a\nabla_a\chi^2}{(\chi\cdot u)^2}.
\end{equation}
It is then most useful to normalize by defining
\begin{equation}
\label{E:theta-n}
\kappa_\mathrm{expansion} = \left.\left\{ (\chi\cdot u) \; \theta_{2d}\right\}\right|_{H},
\end{equation}
which in the case of standard general relativity automatically implies 
\begin{equation}
\kappa_\mathrm{expansion}= \kappa_\mathrm{normal}. 
\end{equation}
This notion of surface gravity is explicitly constructed so that $\kappa_\mathrm{expansion} =  \kappa_\mathrm{normal}$, and hence, in this case, is $\kappa_\mathrm{inaffinity}$.
This derivation relies on the construction of a geodesic congruence that is invariant under the flow of a Killing vector, and so cannot, 
without suitable alterations, be extended to non-Killing horizons which might be present in modified gravity or analogue spacetimes.

\item
Finally, a seventh notion of surface gravity can be based on Euclidean continuation (Wick rotation), and demanding the elimination of the deficit angle at what used to be the horizon in Lorentzian signature. 
This construction of $\kappa_\mathrm{Euclidean}$ is extremely delicate, implicitly requiring constancy of the surface gravity over the horizon (and so implicitly appealing to the rigidity theorems) to even make sense --- but when it works this Euclideanization procedure has the virtue that it automatically forces all quantum fields into an equilibrium thermal bath at the Hawking temperature $k T_H = \hbar \kappa_\mathrm{Euclidean}/2\pi$. 
This procedure works best for static spacetimes, and is already somewhat delicate for stationary non-static spacetimes. 
We will not explore this particular approach any further in the current article.

\end{itemize}
While all of these notions of surface gravity are degenerate in the case of Killing horizons, the situation for non-Killing horizons is much more complex.  
\begin{itemize}
\item 
In standard general relativity it is a well-known result that the surface gravity is constant over the event horizon. 
This result can be proven without recourse to the field equations \emph{if} the horizon is assumed to be Killing~\cite{Racz:1995nh}, but for modified gravity (with field equations that differ from the Einstein equations) one may encounter non-Killing horizons. 
Alternatively, in standard general relativity,  constancy of the surface gravity can be proved using stationarity, the Einstein field equations, and the dominant energy condition for matter~\cite{Hawking:1971vc}. 
(However, note that the dominant energy condition is known to be violated by vacuum polarization effects~\cite{twilight}.) 
In short, this result strongly hinges on the classical equations of motion, and as such, we have no reason to believe this will hold for modified gravity or in analogue spacetime scenarios. 
\item
As a first step beyond standard general relativity, note that even in the case of conformal Killing horizons four of the definitions given in section \ref{standardcase} (inaffinity, normal, generator, tension) do \emph{not} generically coincide. 
This case was considered by Jacobson and Kang~\cite{Jacobson-Kang}. 
The key point is that Jacobson and Kang distinguish several slightly different notions of surface gravity, all of which happen to coincide for Killing horizons (see also~\cite{Nielsen:2012xu, Devecioglu:2011yi}).

The key result (from our current perspective) can be summarized as follows: 
For a conformal Killing vector by definition one has
\begin{equation}\label{conformalkilling}
2\nabla_{(a}\chi_{b)}= \L_\chi g_{ab}= 2F\, g_{ab}.
\end{equation}
Then the relationship between the various surface gravities defined above is
\begin{equation}
\kappa_\mathrm{normal}=\kappa_\mathrm{inaffinity}-2F=\kappa_\mathrm{generator}-F,
\end{equation}
where we have altered their notation to correspond to ours. 
Only one of the definitions can be a true conformal invariant, which they find to be $\kappa_\mathrm{normal}$, while the others will at best be conformally invariant only for those conformal transformations that are constant on the horizon. 
Furthermore $\kappa_\mathrm{tension}$ will be invariant for this special class of transformations, but loses its interpretation for more general conformal transformations.
\end{itemize}
These results, in and of themselves, already provide a clear warning against unrestrictedly interchanging the definitions of surface gravity when working in non-general relativity contexts. 

We shall now discuss two explicit examples of stationary but non-Killing horizons --- one based on the ``analogue spacetime'' programme, and the other on ``universal horizons''.

%------------------------------------------------------------------------------------------------------------------------------------------
%------------------------------------------------------------------------------------------------------------------------------------------
%------------------------------------------------------------------------------------------------------------------------------------------
\section{Analogue spacetimes}\label{adefs}
%------------------------------------------------------------------------------------------------------------------------------------------

In recent years there has been an explosion of interest in the topic of analogue gravity, (more precisely, analogue spacetimes), in part because that framework provides potential for laboratory experiments on some aspects of gravitation~\cite{Visser:1993ub,  Visser:1997ux, Barcelo:2005fc, Visser:2012vh}. 
Theoretically, analogue gravity provides an emergent ``gravitational'' system for which we know the UV physics. As such it has been interesting in shedding light on such issues as the transplanckian problem. 
Additionally, it provides a fascinating test-bed for gaining a deeper understanding of which aspects of gravitation are unique to general relativity or other simple theories of gravitation, which features depend on the field equations, and which are generic geometrical features. 
For a thorough review of analogue gravity see~\cite{Barcelo:2005fc}, and for a shorter introduction see~\cite{Visser:2012vh}.
The simplest model to consider for analogue spacetime is acoustic waves in a fluid system, a model which is extensively developed in section (2) of~\cite{Barcelo:2005fc}. 
For an earlier introduction to some of the features of the scenario considered here see~\cite{Visser:1997ux}. 

%------------------------------------------------------------------------------------------------------------------------------------------
\subsection{Metric}
%------------------------------------------------------------------------------------------------------------------------------------------
We will temporarily restrict ourselves to the case of non-relativistic acoustics in the limit of geometrical acoustics. 
We can write the metric as
\begin{equation}
g_{ab} = \Omega^2 \left[\begin{array}{c|c} -(c_s^2-v^2) & -v_j \\   \hline -v_i & \delta_{ij} \end{array} \right],
\end{equation}
where (for now) the quantities $v_i$ and $c_s$ are position (but not time) dependent. 
The corresponding inverse metric is:
\begin{equation}
g^{ab} = \Omega^{-2} \left[\begin{array}{c|c} -1/c_s^2 & -v^j/c_s^2 \\   \hline -v^i/c_s^2 & \delta^{ij} - v^i v^j / c_s^2\end{array} \right].
\end{equation}
Equivalently, the line element is given by
\begin{equation}
\d s^2 = \Omega^2 \left(-c^2_s\d t^2+ (\d x^i -v^i \d t)(\d x^j -v^j \d t)\delta_{ij} \right).
\end{equation}
For convenience also set
\begin{equation}
\tilde g_{ab} = \left[\begin{array}{c|c} -(c_s^2-v^2) & -v_j \\   \hline -v_i & \delta_{ij} \end{array} \right]; \qquad 
\tilde g^{ab} = \left[\begin{array}{c|c} -1/c_s^2 & -v^j/c_s^2 \\   \hline -v^i/c_s^2 & \delta^{ij} - v^i v^j / c_s^2\end{array} \right].
\end{equation}
Note that indices on $v$ are raised and lowered using $\delta^{ij}$ and $\delta_{ij}$.

%------------------------------------------------------------------------------------------------------------------------------------------
\subsection{Horizon}
%------------------------------------------------------------------------------------------------------------------------------------------
Because of the definition of event horizon in terms of phonons (null geodesics) that cannot escape
the acoustic black hole, the event horizon is automatically a null surface, and the generators of
the event horizon are automatically null geodesics.

Stationary horizons are surfaces, located for definiteness at some $f(\x)=0$, that are defined by the 3-dimensional spatial condition 
\begin{equation}
\vec\nabla f \cdot \v = c_s \; \Vert \vec\nabla f\Vert .
\end{equation}
That is, on a horizon the \emph{normal component} of the fluid velocity equals the speed of sound, thereby either trapping or anti-trapping the acoustic excitations (resulting in black holes or white holes). 

On the horizon we have $(\vec\nabla f \cdot \v)^2 = c_s^2 \; \Vert \vec\nabla f\Vert ^2$, which we can rewrite in 3-dimensional form as $g^{ij} \, \partial_i f \, \partial_j f = 0$, (that is, $[\delta^{ij}-v^iv^j/c_s^2] \, \partial_i f \, \partial_j f = 0$).  
Since the conformation, and location, of the horizon is time independent this statement can be bootstrapped to 3+1 dimensions to see that \emph{on the horizon}
\begin{equation}
g^{ab} \; \nabla_a f \; \nabla_b f = 0. 
\end{equation}
That is, the 4-vector $\nabla f$ is null on the horizon. In fact, on the horizon, where in terms of the (inward-pointing) 3-normal $\n$ we can decompose $\v_H = c_H \; \n +\v_\parallel$ (where the subscript $H$ indicates on-horizon), we can furthermore write
\begin{equation}
 \left(g^{ab} \; \nabla_b f\right)_H  =  {\Vert \vec\nabla f\Vert \over \Omega^2_H \; c_H} \; \left( 1; \v_\parallel\right)_H.
\end{equation}
That is, not only is the 4-vector $\nabla f$ null on the horizon, it is also a 4-tangent to the horizon (note this means we can always apply the Frobenius theorem) --- so, as in general relativity, the horizon is ruled by a set of null curves. 
Furthermore, extending the 3-normal $\n$ to a region surrounding the horizon (for instance by taking $\n = \vec\nabla f/ \Vert \vec\nabla f\Vert $) we can quite generally write
$\v = v_\perp \; \n + \v_\parallel$.  
Then away from the horizon
\begin{equation}
g^{ab} \; \nabla_a f \; \nabla_b f  =    { (c_s^2 - v_\perp^2)\;  \Vert \vec\nabla f\Vert ^2  \over \Omega^2 \; c_s^2}.
\end{equation}
That is, the 4-vector $\nabla f$ is spacelike outside the horizon, null on the horizon, and timelike inside the horizon.

%------------------------------------------------------------------------------------------------------------------------------------------
\subsection{ZAMOs} 
%------------------------------------------------------------------------------------------------------------------------------------------
A rotating analogue black hole, (to be more precise: an analogue black hole where the fluid velocity is not 3-orthogonal to the horizon),  need not be equipped with the same Killing vectors as the Kerr black hole. 
(In particular, the usual theorems whereby stationarity implies axial symmetry need no longer apply.) 
To attempt to generalize the constructions in Wald~\cite{Wald}, we want a natural vector that is timelike outside, spacelike inside, and null on the horizon. 
For this we will consider a vector describing an observer similar to a ZAMO (zero angular momentum observer, see for instance~\cite{raine-thomas}). 
To capture a suitable notion of ``comoving with the horizon'' let us define
\begin{equation}
Z^a = (1; \; \v_\parallel); \qquad Z_a = -\Omega^2  ( c_s^2-v_\perp^2; \; v_\perp \n).
\end{equation}
Then we have $g_{ab} \; Z^a \; Z^b=  - \Omega^2 (c_s^2- v_\perp^2)$, which is null on the horizon.
Furthermore $Z^a \partial_a f \equiv 0$, so these vector fields $Z^a$ foliate the constant-$f$ surfaces, $f(\x)=C$, and in particular foliate the horizon at $f(\x)=0$. 
In the current context the vector $Z^a$ is the closest we can get to a horizon-foliating Killing vector; it is at least horizon-foliating, even if it is not necessarily Killing.
For later convenience, we also define 
\begin{equation}
\tilde Z^a = (1; \; \v_\parallel)= z^a; \qquad \tilde Z_a = - ( c_s^2-v_\perp^2; \; v_\perp \n)=\frac{Z_a}{\Omega^2}.
\end{equation}

%------------------------------------------------------------------------------------------------------------------------------------------
\subsection{The on-horizon Lie derivative}
%------------------------------------------------------------------------------------------------------------------------------------------

Note the Lie derivative
\begin{equation}
(\mathcal{L}_Z g)_{ab} = Z_{a;b} + Z_{b;a} = Z^c{}_{,a} g_{cb} + Z^c{}_{,b} g_{ca}  + Z^c{} \partial_c  g_{ab},
\end{equation}
evaluates to
\begin{equation}\label{liederiv}
(\mathcal{L}_Z g)_{ab} = \Omega^2 (\mathcal{L}_{\tilde Z} \tilde g)_{ab} + 2 (\v_\parallel\cdot \vec\nabla \ln\Omega) g_{ab}.
\end{equation}
Explicitly
\begin{equation}
(\mathcal{L}_Z g)_{ab} =
\Omega^2 \left[ \begin{array}{c|c}  - \v_\parallel\cdot\vec\nabla (c^2-v^2)  & - v_\parallel{}^k{}_{,i} v_k - v_\parallel{}^k \partial_k v^i\\
\hline
- v_\parallel{}^k{}_{,j} v_k - v_\parallel{}^k \partial_k v^j &  v_{\parallel\, i,j} + v_{\parallel\, j,i} \end{array} \right]
 + 2 (\v_\parallel\cdot \vec\nabla \ln\Omega) g_{ab}.
\end{equation}
It is the fact that this quantity is non-vanishing that makes the horizon non-Killing. 
The $(\v_\parallel\cdot \vec\nabla \ln\Omega)$ term is just a conformal Killing contribution, hence more or less ``trivial'' (apply the Jacobson--Kang~\cite{Jacobson-Kang} argument). Now, on-horizon,
\begin{equation}
(\mathcal{L}_Z g)^H_{ab} =
\Omega^2 \left. \left[ \begin{array}{c|c}  \v_\parallel\cdot\vec\nabla (v_\parallel^2)    & - v_\parallel{}^k{}_{,i} v_k - v_\parallel{}^k \partial_k v_i\\
\hline
- v_\parallel{}^k{}_{,j} v_k - v_\parallel{}^k \partial_k v_j &  v_{\parallel\, i,j} + v_{\parallel\, j,i} \end{array} \right]\right|_{H}
 + \left. 2 (\v_\parallel\cdot \vec\nabla \ln\Omega) g_{ab} \right|_{H}.
\end{equation}
We can write this in terms of the 3-d spatial Lie derivative (with respect to $\v_\parallel$) as
\begin{equation}
(\mathcal{L}_Z g)^H_{ab} =
\Omega^2 \left. \left[ \begin{array}{c|c}  \L_{v_\parallel} (v_\parallel^2)    & - \L_{v_\parallel} v_i\\
\hline
- \L_{v_\parallel}v_j & + \L_{v_{\parallel}} \delta_{ij} \end{array} \right]\right|_H
 + \left. 2 (\v_\parallel\cdot \vec\nabla \ln\Omega) g_{ab} \right|_H .
\end{equation}
This makes it obvious that it is the in-horizon symmetries (or lack thereof) which governs whether or not the horizon is Killing. 
From this perspective, the key reason for the degeneracy of surface horizon definitions in general relativity is that the field equations impose symmetries \emph{on horizon}.
Comparing equation (\ref{liederiv}) to equation (\ref{conformalkilling}) we can clearly see how our how our results in the next section correspond to and extend those of Jacobson and Kang~\cite{Jacobson-Kang}.

%------------------------------------------------------------------------------------------------------------------------------------------
\subsection{Surface gravities}
%------------------------------------------------------------------------------------------------------------------------------------------

We shall now evaluate the various definitions of surface gravity by explicit calculation.

%------------------------------------------------------------------------------------------------------------------------------------------
\subsubsection{Geodesic peeling}
%------------------------------------------------------------------------------------------------------------------------------------------

In the spherically symmetric case, we previously considered the peeling properties of \emph{radial} null geodesics. 
In contrast, here we want \emph{corotating} null geodesics, that is, outgoing null geodesics that are as close as possible to ZAMOs. Furthermore, as these geodesics emerge from the region near the horizon, their 3-velocity will have a normal component, the ``speed'' with which it is escaping ``vertically''. 
That is, take
\begin{equation}
k^a = (1, -\dot h \n + \v_\parallel);
\end{equation}
here $h$ denotes a normal height above the horizon, and dot indicates a time derivative. 
The null condition,
\begin{equation}
g_{ab} k^a k^b=0,
\end{equation} 
yields
\begin{equation}
-(c_s^2-v^2) - 2(-\dot h v_\perp + v_\parallel^2) + \dot h^2 + v_\parallel^2 = 
- c_s^2 + (\dot h + v_\perp)^2 = 0.
\end{equation} 
Thence we have the very simple and physically plausible result
\begin{equation}
\dot h =  \pm c_s - v_\perp.
\end{equation}
For those null curves that are just escaping, near the horizon we have
\begin{equation}
\left.\dot h = c_s - v_\perp \approx - {\partial(c_s-v_\perp)\over\partial n}\right|_H \; h.
\end{equation}
(Remember $\n$ is inward pointing.) 
Let us define:
\begin{equation}
\left.\kappa_\mathrm{peeling} = -{\partial(c_s-v_\perp)\over\partial n}\right|_H =  c_H \; {\partial M_\perp\over\partial n},
\end{equation}
where $M_\perp = v_\perp/c_s$ is the transverse Mach number. 
Note this quantity $\kappa_\mathrm{peeling}$ is manifestly conformally invariant. 
Also  $\kappa_\mathrm{peeling}$ is \emph{not} necessarily constant over the horizon; the steepness of the the Mach number is \emph{not} constrained automatically to be the same everywhere along the horizon.
Then
\begin{equation}
h \approx h_*\; \exp({\kappa_\mathrm{peeling}[t-t_*]}).
\end{equation}
This is clearly related to the peeling off ($e$-folding) properties of escaping null curves near the horizon. 

%------------------------------------------------------------------------------------------------------------------------------------------
\subsubsection{Null gradient normal to horizon} 
%------------------------------------------------------------------------------------------------------------------------------------------
(It is best to consider this particular notion slightly ``out of order'', as $\kappa_\mathrm{normal}$ will prove useful when discussing $\kappa_\mathrm{inaffinity}$.) 
The gradient normal definition of surface gravity always works for acoustic horizons as we have defined them above, as on the horizon $Z^b Z_b =0$, and so its gradient is normal to the horizon. 
\emph{If we have already decided that the horizon is a null surface}, then its null normal must lie in the horizon, and so be proportional to $Z$. 
Then there must be a scalar $\kappa_\mathrm{normal}$ such that: 
\begin{equation}\label{normalkappa}
\nabla_a (Z^b Z_b) = - 2 \kappa_\mathrm{normal}\; Z_a.
\end{equation}
Equivalently
\begin{equation}\label{normalkappa-bis}
Z^b \nabla_a Z_b = - \kappa_\mathrm{normal} \; Z_a.
\end{equation}
But by explicit computation we now see
\begin{eqnarray}
\left.\nabla_a (Z^b Z_b)\right|_H &=& \nabla_a \left[ -\Omega^2 (c_s^2- v_\perp^2)\right] 
= - 2\Omega^2 \left( 0 ; \; c_s \nabla_i (c_s-v_\perp)\right) 
\nonumber
\\ 
&=& - 2\Omega^2 c_s {\partial(c_s-v_\perp)\over\partial n}\left( 0; \; \n\right)
= 2 {\partial(c_s-v_\perp)\over\partial n} \; Z_{a|H},
\end{eqnarray}
where
\begin{equation}
Z_{a|H} =  - \Omega^2 c_H (0;\; \n).
\end{equation}
Therefore with  this definition:
\begin{equation}
\kappa_\mathrm{normal} = -{\partial(c_s-v_\perp)\over\partial n} =  c_H \; {\partial M_\perp\over\partial n} = \kappa_\mathrm{peeling}.
\end{equation}
So we explicitly see that the peeling and normal gradient notions of surface gravity are still degenerate for acoustic horizons.

%------------------------------------------------------------------------------------------------------------------------------------------
\subsubsection{Inaffinity}
%------------------------------------------------------------------------------------------------------------------------------------------

Now consider the inaffinity definition of surface gravity. We would like to be able to write
\begin{equation}
Z^b \nabla_b Z_a =  \kappa_\mathrm{inaffinity} \; Z_a.
\end{equation}
Our first problem is that, although $Z^a$ is null, we have no  \emph{a priori} reason to expect $Z^b \nabla_b Z_a $ to be null, despite being automatically orthogonal to $Z^a$. 
We need to show that our horizon is what we will term as ``geodesic'', that is, foliated by null geodesics. 
Note that (on horizon) we \emph{always} have:
\begin{eqnarray}\label{geodesichorizon}
Z^b \nabla_b Z^a &=& Z^b (\nabla_b Z^a +\nabla^a Z_b) - {1\over2}\nabla^a (Z^b Z_b)
\nonumber \\
 &=& (\L_Z g)^a{}_b Z^b + \kappa_\mathrm{normal} Z^a.  
\end{eqnarray}
(The occurrence of the quantity $(\L_Z g)^a{}_b$ above is the explicit signal of a possible non-Killing horizon, and the reason we discussed and evaluated this quantity previously.)
On the horizon $Z^a$ is guaranteed null; both $Z^b \nabla_b Z^a$ and $(\L_Z g)^a{}_b Z^ b$ are guaranteed to be orthogonal to $Z$, but without further assumptions we cannot guarantee that they are null.
\emph{If (for now) we simply assume the horizon is geodesic}, that is, foliated by null geodesics, then
\begin{equation}
Z^b \nabla_b Z^a = \kappa_\mathrm{inaffinity} \; Z^a,
\end{equation}
and then
\begin{equation}
(\L_Z g)^a{}_b Z^ b = (\kappa_\mathrm{inaffinity}-\kappa_\mathrm{normal} ) \; Z^a = \Delta \kappa \; Z^a.
\end{equation}
Note the condition $(\L_Z g)^a{}_b Z^ b = \Delta \kappa \; Z^a$ is equivalent to demanding
\begin{equation}
(\L_Z g)_{ab} = \Delta \kappa \; g_{ab} + \zeta \; Z_a Z_b + \xi\; P^\perp_{ab}.
\end{equation}
Here 
\begin{equation}
P^\perp_{ab}=g_{ab}+{Z_aZ_b\over\Vert Z\Vert^2}.
\end{equation}

\bigskip
\noindent
This construction defines a hierarchy of possible horizons: 
\begin{itemize}
\item 
Killing ($\L_Z g=0$, the standard GR case); 
\item
conformally Killing ($\Delta\kappa\neq0$, $\zeta=\xi=0$, the Jacobson--Kang generalization); 
\item
``Kerr--Schild-like''  ($\Delta\kappa\neq0$, $\zeta\neq0$, $\xi=0$); 
\item general geodesic  ($\Delta\kappa\neq0$, $\zeta\neq 0$, $\xi\neq0$, our current case). 
\end{itemize}
The horizons we have termed ``Kerr--Schild-like'', where $(\L_Z g)_{ab}$ is of Kerr--Schild form on the horizon, have not to the best of our knowledge, been separately studied. 
We will now \emph{prove} that all the acoustic horizons we are considering are geodesic horizons, a fact that will also be used in the analysis of the next definition ($\kappa_\mathrm{generator}$).

We see from equation (\ref{geodesichorizon}) that
\begin{eqnarray}
\label{geodesichorizon2}
Z^b \nabla_b Z^a &=& Z^b (\nabla_b Z^a -\nabla^a Z_b) + {1\over2}\nabla^a (Z^b Z_b)
\nonumber \\
 &=&  Z^b (\nabla_b Z^a -\nabla^a Z_b)  - \kappa_\mathrm{normal} Z^a.  
\end{eqnarray}
Thus the horizon is geodesic iff (on the horizon)
\begin{equation}
Z^b (\nabla_b Z_a -\nabla_a Z_b)  \propto Z_a.
\end{equation}
Recall the definitions of $Z^a$, $g_{ab}$, $\tilde Z^a$ and $\tilde g_{ab}$ given in section (\ref{adefs}). 
We note
 \begin{equation}
  \nabla_{[a} Z_{b]} =  \Omega^2 \nabla_{[a} \tilde Z_{b]} + 2\nabla_{[a} \ln\Omega \; Z_{b]},
\end{equation}
where on the horizon
  \begin{equation}
  (\nabla_{[a} \tilde Z_{b]})_H = 
  -\left[\begin{array}{cc} 0 & c_H \,\kappa_\mathrm{normal} \, n_j \\ 
  - c_H \,\kappa_\mathrm{normal} \,n_i&  (v_\perp n_{[i})_{,j]}\end{array}\right].
  \end{equation}
But by definition we have $n_i = \partial_i f /\Vert \partial f\Vert $, so
\begin{equation}
(v_\perp n_{[i})_{,j]} = -(v_\perp/\Vert \partial f\Vert )_{[,i}  f_{,j]} = -\Vert \partial f\Vert  (v_\perp/\Vert \partial f\Vert )_{[,i} n_{j]} =c_H \tilde s_{[i} n_{j]},
\end{equation}
where we now define
\begin{equation}
\tilde s_i \equiv  - {\Vert \partial f\Vert \over c_H} \; (v_\perp/\Vert \partial f\Vert )_{,i} = - {v_{\perp,i}\over c_H} + \partial_i \ln \Vert \partial f\Vert ,
\end{equation}
with dimensions $[\tilde s] = {1/[L]}$. Therefore
 \begin{equation}
  (\nabla_{[a} \tilde Z_{b]})_H = \left[\begin{array}{cc} 0 & -c_H  \,\kappa_\mathrm{normal} \,  n_j \\  
  c_H  \,\kappa_\mathrm{normal} \,  n_i &  c_H n_{[i} \tilde s_{j]}\end{array}\right].
  \end{equation}
Now defining $\tilde S_a=(2\kappa_\mathrm{normal}, \tilde s_i)$,  we see that on the horizon
 \begin{equation}
  (\nabla_{[a} \tilde Z_{b]})_H = \tilde Z_{[a} \tilde S_{b]}.
 \end{equation}
 Thence, defining $S_a = \tilde S_a  - 2\nabla_{a} \ln\Omega $ we see that on the horizon
 \begin{equation}
 (\nabla_b Z_a -\nabla_a Z_b)_H = Z_a S_b - S_a Z_b.
 \end{equation}
 But then
 \begin{equation}
 (\nabla_b Z_a -\nabla_a Z_b)_H Z^b = Z_a (S_b Z^b) =  (2\kappa_\mathrm{normal} + \v_\parallel \cdot \s) Z_a.
 \end{equation}
This observation is already enough to guarantee that the horizon is geodesic.

But now that we have shown that the horizon is geodesic, it follows immediately that we have the even stronger statement:
\begin{equation}
\kappa_\mathrm{inaffinity}  =   (S_b Z^b) - \kappa_\mathrm{normal} = \kappa_\mathrm{normal} + \v_\parallel \cdot \s = \kappa_\mathrm{normal} - 2\v_\parallel\cdot\nabla \ln\Omega + \v_\parallel \cdot \tilde{\s}. 
\end{equation}
But now
\begin{equation}
 \v_\parallel \cdot \tilde{\s} =  -{\v_\parallel \cdot \nabla v_\perp\over c_H} + { \v_\parallel \cdot \nabla \ln\Vert \partial f\Vert } 
 =  -\v_\parallel \cdot \nabla \ln c_H +  \v_\parallel \cdot \nabla \ln\Vert \partial f\Vert .
\end{equation}
Furthermore
\begin{eqnarray}
\v_\parallel \cdot \nabla \ln\Vert \partial f\Vert  &=& {1\over2} \v_\parallel \cdot \nabla \ln[\Vert \partial f\Vert ^2] 
= {v_\parallel^i \; f_{,ij} \; f_j \over \Vert \partial f\Vert ^2} 
= - {v_\parallel^i{}_{,j} \; f_{,i} \; f_j \over \Vert \partial f\Vert ^2} 
\nonumber\\
&=&  -v_\parallel{}^{i,j} n_i n_j =  -v_\parallel{}^{(i,j)} \; n_i n_j.
\end{eqnarray}
Pulling it all together
\begin{equation}
\kappa_\mathrm{inaffinity}  = \kappa_\mathrm{normal} - 2\v_\parallel \cdot\nabla \ln\Omega - \v_\parallel \cdot \nabla \ln c_H -v_\parallel{}^{(i,j)} \; n_i n_j.
\end{equation}
The last term is an internal horizon \emph{shear}. 
This quantity $\kappa_\mathrm{inaffinity}$ is manifestly \emph{not} a conformal invariant. 
One can also express this as
\begin{equation}
\kappa_\mathrm{inaffinity}  = \kappa_\mathrm{normal} - \v_\parallel \cdot \nabla \ln [c_H \Omega^2] -v_\parallel{}^{(i,j)} \; n_i n_j. 
\end{equation}
This is consistent with the Jacobson--Kang analysis, as for them, automatically, the in-horizon shear is taken to be zero.

%------------------------------------------------------------------------------------------------------------------------------------------
\subsubsection{Generator-based} 
%------------------------------------------------------------------------------------------------------------------------------------------
We shall define
\begin{equation}
\kappa_\mathrm{generator}^2 = -{1\over2} (\nabla^{[a} Z^{b]})_H (\nabla_{[a} Z_{b]})_H.
\end{equation}
We can always \emph{define} this quantity into existence, the question is how does it relate to the previous two definitions? 

We have already shown that at the analogue horizon
 \begin{equation}
 (\nabla_b Z_a -\nabla_a Z_b)_H = Z_a S_b - S_a Z_b.
 \end{equation}
 But then
\begin{equation}
(\nabla_b Z_a -\nabla_a Z_b)_H (\nabla^b Z^a -\nabla^a Z^b)_H = - 2(S_a Z^a)^2.
\end{equation}
Therefore
\begin{eqnarray}
\kappa_\mathrm{generator}^2 &=& -{1\over2} (\nabla^{[a} Z^{b]})_H (\nabla_{[a} Z_{b]})_H \nonumber\\
&=&  -{1\over8}  (\nabla^b Z^a -\nabla^a Z^b)_H (\nabla_b Z_a -\nabla_a Z_b)_H \nonumber\\
&=& \; \; \, {1\over4} (S_a Z^a)^2,
\end{eqnarray}
and so
\begin{equation}
\kappa_\mathrm{generator} = {1\over2} (S_a Z^a) = {\kappa_\mathrm{normal}+\kappa_\mathrm{inaffinity}\over2}.
\end{equation}
Pulling it all together we see
\begin{equation}
\kappa_\mathrm{generator}  = \kappa_\mathrm{normal} - \v_\parallel \cdot\nabla \ln\Omega - {1\over2}\v_\parallel \cdot \nabla \ln c_H -{1\over2}v_\parallel{}^{(i,j)} \; n_i n_j.
\end{equation}
Alternatively,
\begin{equation}
\kappa_\mathrm{generator}  = \kappa_\mathrm{normal}- {1\over2}\v_\parallel \cdot \nabla \ln [c_H\Omega^2] -{1\over2}v_\parallel{}^{(i,j)} \; n_i n_j.
\end{equation}
This quantity is manifestly \emph{not} conformally invariant.

%------------------------------------------------------------------------------------------------------------------------------------------
\subsubsection{Tension in a rope}
%------------------------------------------------------------------------------------------------------------------------------------------

There is a nice argument leading to a tidy physical interpretation of the surface gravity in terms of tension in an ideal massless rope held at infinity. 
In the current context we would want to evaluate
\begin{equation}
\kappa_\mathrm{tension} = \lim_H \sqrt{-Z^2}\, \Vert A\Vert ,
\end{equation}
with $A$ the magnitude of the 4-acceleration of the integral curves of $Z^a$. Define
\begin{equation}
V^a = {Z^a\over\sqrt{-Z^b Z_b}}, \qquad A^a = V^b \nabla_b V^a,
\end{equation}
as the velocity and acceleration of an orbit of $Z^a$. Now using
\begin{equation}
Z^a = \sqrt{-Z^b Z_b}\; V^a,
\end{equation}
we see
\begin{equation}
Z_b \nabla^b Z^c =  (-Z^b Z_b) A^c + {1\over2} {Z^b \nabla_b (-Z^2)\over (-Z^2)} Z^c. 
\end{equation}
Then working outside the horizon, where $A$ and $Z$ are 4-perpendicular, and $Z$ is timelike while $A$ is spacelike,  we have
\begin{equation}
  (-Z^b Z_b) \Vert A^c\Vert ^2 = { \Vert Z_b \nabla^b Z^c \Vert ^2 \over (-Z^2) }+  {1\over4} {[Z^b \nabla_b \ln (-Z^2)]^2}.
\end{equation}
Now, as we approach the horizon
\begin{equation}
{\Vert Z_b \nabla^b Z^c \Vert ^2 \over (-Z^2) }  \to  {0\over0}.
\end{equation}
Since this is indeterminate it is useful to consider
\begin{eqnarray}
{ \nabla_a\Vert Z_b \nabla^b Z^c \Vert ^2 \over \nabla_a (-Z^2) } 
&\to& 
{ 2(Z_b \nabla^b Z^c)\nabla_a(Z_b \nabla^b Z^c)  \over -2\kappa_\mathrm{normal} Z_a} \nonumber\\
&=& 
{ (\kappa_\mathrm{inaffinity} Z_c)\nabla_a(Z_b \nabla^b Z^c)  \over -\kappa_\mathrm{normal} Z_a} \nonumber\\
&=& 
{ \kappa_\mathrm{inaffinity} (\nabla_a( Z_c Z_b \nabla^b Z^c) - (\nabla_a Z_c)  (Z_b \nabla^b Z^c)) \over -\kappa_\mathrm{normal} Z_a} \nonumber\\
&=& 
{ \kappa_\mathrm{inaffinity} (\nabla_a(0) - (\nabla_a Z_c)  (\kappa_\mathrm{inaffinity} Z^c)) \over -\kappa_\mathrm{normal} Z_a} \nonumber\\
&=& 
{ -\kappa_\mathrm{inaffinity}^2 (\nabla_a Z_c) Z^c \over -\kappa_\mathrm{normal} Z_a} \nonumber\\
&=& 
\kappa_\mathrm{inaffinity}^2 \; \left({ \kappa_\mathrm{normal} Z_a \over \kappa_\mathrm{normal} Z_a}\right) \nonumber\\
&=& 
\kappa_\mathrm{inaffinity}^2.
\end{eqnarray}
So by the l'Hospital rule:
\begin{equation}
\lim_H \left\{  { \Vert Z_b \nabla^b Z^c \Vert ^2 \over \Vert Z\Vert ^2) } \right\} = \kappa_\mathrm{inaffinity}^2.
\end{equation}
Furthermore, as we approach the horizon
\begin{equation}
Z^2 = -2\Omega^2(c_s^2-v_\perp^2) \approx  -\Omega^2 c_H \kappa_\mathrm{normal} \times (\hbox{normal 3-distance to horizon}).
\end{equation}
So 
\begin{equation}
\lim_H \left(Z^d \nabla_d \ln(-Z^2) \right)  = \v_\parallel\cdot\nabla\ln[\Omega^2 c_H \kappa_\mathrm{normal}] .
\end{equation}
(Remember that for an acoustic horizon there is no need to believe in a zeroth law, there is no need for $\kappa_\mathrm{normal}$ to be constant over the horizon).
Pulling everything together
\begin{equation}
\kappa_\mathrm{tension}^2 = \lim_H \{(-Z^2) \Vert A^c\Vert ^2\} =  
\kappa_\mathrm{inaffinity}^2 +  {1\over4} \left(\v_\parallel\cdot\nabla\ln[\Omega^2 c_H \kappa_\mathrm{normal}] \right)^2.
\end{equation}
That is:
\begin{equation}
\kappa_\mathrm{tension} = \sqrt{\kappa_\mathrm{inaffinity}^2 +  {\textstyle{1\over4}} \left(\v_\parallel\cdot\nabla\ln[\Omega^2 c_H \kappa_\mathrm{normal}] \right)^2}.
\end{equation}
Now using
\begin{equation}
\kappa_\mathrm{inaffinity}  = \kappa_\mathrm{normal} - \v_\parallel \cdot \nabla \ln [c_H \Omega^2] -v_\parallel{}^{(i,j)} \; n_i n_j ,
\end{equation}
we have
\begin{equation}
\kappa_\mathrm{tension} = \sqrt{\left(\kappa_\mathrm{normal} - \v_\parallel \cdot \nabla \ln [c_H \Omega^2] -v_\parallel{}^{(i,j)} \; n_i n_j \right)^2 
+  {\textstyle{1\over4}} \left(\v_\parallel\cdot\nabla\ln[\Omega^2 c_H \kappa_\mathrm{normal}] \right)^2}.
\end{equation}
This quantity is manifestly \emph{not} a conformal invariant.

%------------------------------------------------------------------------------------------------------------------------------------------
\subsubsection{2-d expansion}
%------------------------------------------------------------------------------------------------------------------------------------------

Finally, there is a recent argument by Jacobson and Parentani~\cite{Jacobson:2008cx}, relating the surface gravity to
the expansion of a suitably defined congruence of timelike geodesics normal to the horizon~\cite{Jacobson:2008cx}. See earlier discussion and equations (\ref{E:theta-0})--(\ref{E:theta-n}).
The key point here, is once again this equality relies on the existence of an appropriate geodesic congruence invariant under the flow of a Killing (or Killing-like) vector, and so cannot be applied blindly to modified gravity or analogue gravity scenarios.

For an acoustic horizon we would want to pick a congruence dragged by $Z^a$,
\begin{equation}
Z^a \nabla_a u^b=u^a\nabla_a Z^b.
\end{equation}
\emph{If} it is possible to construct such a congruence, then from equation (\ref{normalkappa}), we know that 
\begin{equation}
\theta_{2d}  =\frac{(u\cdot Z) \kappa_\mathrm{normal}}{(Z\cdot u)^2}.
\end{equation}
And hence, now for an acoustic horizon,
\begin{equation}
\kappa_\mathrm{expansion}=(u \cdot Z)\theta_{2d}= \kappa_\mathrm{normal}. 
\end{equation}

%------------------------------------------------------------------------------------------------------------------------------------------
\subsubsection{Summary}
%------------------------------------------------------------------------------------------------------------------------------------------

For an acoustic horizon we generically have
\begin{equation}
\kappa_\mathrm{normal} = \kappa_\mathrm{peeling} = \kappa_\mathrm{expansion}.
\end{equation}
On the other hand $\kappa_\mathrm{inaffinity}$, $\kappa_\mathrm{generator}$, and $\kappa_\mathrm{tension}$ are generically distinct from each other, and from the preceding three items.

%------------------------------------------------------------------------------------------------------------------------------------------
%------------------------------------------------------------------------------------------------------------------------------------------
%------------------------------------------------------------------------------------------------------------------------------------------
\section{Modified gravity}
%------------------------------------------------------------------------------------------------------------------------------------------
While in the previous section we have been interested in the framework of analogue gravity, the concerns we have are also of vital importance for modified gravity. Some general points to consider:

\begin{itemize}

\item The usual situation, where the final state of a black hole is either static, or stationary and axisymmetric, depends critically on the standard Einstein equations (and ``reasonable'' matter sources). 
This could easily fail in modified gravity.

\item The usual situation, where black hole horizons are Killing horizons,  depends critically on the standard Einstein equations (and ``reasonable'' matter sources), which could easily fail in modified gravity~\cite{HawkingEllis, carter1, carter2, Heusler:1998, Friedrich:1998wq, Robinson:2004, Kodama:2011}.

\item The usual situation, where black holes satisfy the zeroth law (constancy of $\kappa$), depends critically on the ``effective stress energy'', in the sense $G^{ab} \propto T^{ab}_\mathrm{effective}$, satisfying some form of classical energy condition. Again, this could easily fail in modified gravity~\cite{Racz:1995nh,Hawking:1971vc}. 
\end{itemize}
In short, the distinctions between the various surface gravities can also easily become important outside of the analogue spacetime framework. We will work through one specific example within the framework of Lorentz-violating theories to demonstrate this. 

%------------------------------------------------------------------------------------------------------------------------------------------
\subsection{Einstein-aether and Horava--Lifshitz gravity}
%------------------------------------------------------------------------------------------------------------------------------------------

Einstein-aether and Horava--Lifshitz gravity are two theories of gravity which violate Lorentz-invariance.  
Einstein-aether theory (first proposed in~\cite{Gasperini:1987nq,Gasperini:1998eb}, and developed further in~\cite{Jacobson:2000xp}), is general relativity coupled to a dynamical, unit timelike vector.  
Einstein-aether theory was originally constructed  as a mechanism for breaking local Lorentz symmetry yet retaining as many of the other positive characteristics of general relativity as possible. 
In particular it is described by the most general action involving the metric and a unit timelike vector $u^a$ that contains no more than second-order derivatives in the fields and is generally covariant.

Horava gravity, proposed in~\cite{Horava:2009uw}, is another theory with Lorentz violation, in this case motivated by aims to construct a renormalizable model of quantum gravity by giving up Lorentz-invariance, as the ultraviolet behavior can be substantially improved by the addition of terms with higher spatial derivatives to the action. 
Indeed, Horava--Lifshitz gravity is power-counting renormalizable~\cite{Horava:2009uw}. 

A particular variety of Horava--Lifshitz gravity, non-projectable Horava--Lifshitz gravity, in the IR limit, becomes Einstein-aether theory when the aether vector is restricted to be hypersurface orthogonal (note that this is automatically the case for spherically symmetric solutions). 
See~\cite{Jacobson:2010mx, Sotiriou:2010wn}. 

One important feature of Horava--Lifshitz gravity, is that the action for matter will have to include higher order spatial derivatives. 
These dispersion relations can easily lead to situations such that  there is no limiting speed in the theory. 
This, and the notion of a time-defining aether, means that the causal features of the theory are completely different than that of general relativity; for instance, it is not obvious \emph{a priori} that any sort of black hole would exist in such theory.

%------------------------------------------------------------------------------------------------------------------------------------------
\subsection{Universal horizons and their surface gravities}
%------------------------------------------------------------------------------------------------------------------------------------------
It has recently been realized (see~\cite{Berglund:2012fk, Berglund:2012bu, Barausse:2011pu, Blas:2011ni}), that (spherically symmetric) black holes in Lorentz-violating theories do exist, and contain, inside the standard Killing horizon, a new sort of horizon, essentially a surface where to flow forward in time, particles must enter the (spacelike) surface, defined as the surface where $\chi \cdot u=0$. 
This is significant, in that it shows that, even in the case of Horava--Lifshitz gravity, where dispersions relations will remove the causal significance of any horizon for a finite-speed mode, there \emph{is} a notion of a causal boundary in such theories.
From our point of view, these universal horizons are interesting because they provide examples of non-Killing horizons, and furthermore these horizons are \emph{not} null surfaces, unlike the cases we have previously been looking at. 
Relevant questions are:
\begin{itemize}
\item Which of the definitions of surface gravity can be extended to these universal horizons?
\item Are these all identical? If not, how do they differ?
\end{itemize}
These are non-trivial questions, important for questions such as whether or not Hawking radiation exists for such theories, from what surface, and further the wider issues surrounding the thermodynamics of such spacetimes.

%------------------------------------------------------------------------------------------------------------------------------------------
\subsubsection{Generator-based}
%------------------------------------------------------------------------------------------------------------------------------------------

This is the quantity calculated in reference~\cite{Berglund:2012bu}; we reproduce the most salient aspects of the derivation here. 
(We will carefully work through this definition first, as it is the one used in previous literature, and our subsequent constructions rely heavily on this set-up).

Set up a tetrad of unit vectors, the timelike vector given by the aether, $u^a$, then two spacelike vectors $m^a$ and $n^a$, which are mutually orthogonal and lie in the tangent plane of two-spheres, and a spacelike unit vector is provided by the outward-pointing $s^a$ (our radial vector). 
Further, any rank-two tensor can be expanded in terms of the quantities $u_au_b$, $u_{(a}s_{b)}$, $u_{[a}s_{b]}$, $s_as_b$, and $\hat g_{ab}$; where $\hat g_{ab}$ is projection tensor onto the spatial two-sphere surface.

As we have spherical symmetry any physical vector should have components only along $u^a$ and $s^a$. 
Also note the acceleration will only have a component along $s^a$. That is, $a^a=(a\cdot s)s^a$. 
Further note that at the universal horizon $s^a$ is, by definition, parallel to $\chi^a$.
We therefore have the useful relations:
\begin{equation}
\nabla_a u_b = -(a\cdot s)\; u_a s_b + K^{(u)}_{a b}\,; \qquad
K^{(u)}_{a b} = K_0 \; s_a s_b + \frac12{\hat{K}^{(u)}}\;\hat{g}_{a b},
\end{equation}
\begin{equation}
\nabla_a s_b = K_0 \; s_a u_b + K^{(s)}_{a b}\,; \qquad
K^{(s)}_{a b} = -(a\cdot s)\; u_a u_b + \frac12{\hat{K}^{(s)}}\;\hat{g}_{a b}.
\end{equation}
Here $K^{(u)}_{a b}$ is the extrinsic curvature of the hypersurfaces orthogonal to the aether flow $u^a$, while $ K^{(s)}_{a b}$ is the extrinsic curvature of the hypersurfaces orthogonal to $s^a$,  and $\hat{K}^{(u)}$ and $\hat{K}^{(s)}$ are the traces of the extrinsic curvatures of the 2-spheres due to their embeddings in these two hypersurfaces, respectively. Finally $K_0$ is related to the 4-acceleration of the integral curves of $s^a$ by $s^a\nabla_a s_b = K_0 \; u_b$.

Now consider an arbitrary vector of form
\begin{equation}
A_a=-fu_a+hs_a,
\end{equation}
where $f$ and $h$ are arbitrary functions respecting the symmetries of the spacetime, so in particular $A$ is Lie dragged by the Killing vector $\chi$.
By spherical symmetry
\begin{equation}\label{derivphysvector}
\nabla_{\left[a\right.}A_{\left.b\right]}= - Q_A \; u_{\left[a\right.}s_{\left.b\right]},
\end{equation}
(as this is the only possible fully anti-symmetric choice possible within spherical symmetry), with
\begin{equation}\label{generalq}
Q_A=-f(a\cdot s)-s^a\nabla_a f +hK_0+u^a\nabla_a h.
\end{equation}
We have chosen an opposite sign convention to Berglund \emph{et al.}~\cite{Berglund:2012bu} to minimize subsequent sign flips.

Our Killing vector is 
\begin{equation}
\chi^a=-(u\cdot \chi)u^a +(s\cdot \chi)s^a.
\end{equation}
And from the results above, and the Killing equation, we have
\begin{equation}\label{killingderiv}
\nabla_a\chi_b=- \frac{Q_\chi}{2}(u_as_b-s_au_b),
\end{equation}
where now
\begin{eqnarray}
\label{E:tricky}
Q_\chi &=&- (u\cdot \chi)(a\cdot s) +(s\cdot \chi)K_0-s^a\nabla_a(u\cdot \chi)+u^a\nabla_a(s\cdot \chi) \nonumber\\
&=& -2\left\{ (u\cdot \chi)(a\cdot s) -(s\cdot \chi)K_0 \right\}.
\end{eqnarray}
The second equality follows from the fact that for any $A$ respecting the symmetries of the spacetime
\begin{eqnarray}
\nabla_a (A\cdot \chi) 
&=& (\nabla_a\chi^b) A_b + \chi^b \nabla_a A_b 
\nonumber\\
&=&-\chi^b\nabla_b A_a + \chi^b \nabla_a A_b 
\nonumber \\
&=& -Q_A \{ (s \cdot \chi) u_a- (u \cdot \chi) s_a\}.
\end{eqnarray}
Specializing this relation to our case we have
\begin{eqnarray}
s^a\nabla_a(u \cdot \chi)  &=& Q_{u}(u \cdot \chi) = (a\cdot s) \, (u \cdot \chi); \\
u^a\nabla_a(s \cdot \chi)  &=& Q_{s}(s \cdot \chi) = K_0\; (s \cdot \chi).
\end{eqnarray}
Combining these results we obtain the second line of (\ref{E:tricky}).

We can now identify $\kappa_\mathrm{generator}$ with $|Q_\chi|/2$, as given in equations (22) and (23) of reference \cite{Berglund:2012bu}, since, provided (\ref{killingderiv}) holds true, we have
\begin{equation}
\kappa_\mathrm{generator}=\sqrt{-\frac{1}{2}(\nabla_a\chi_b)(\nabla^a\chi^b)} = \frac{|Q_\chi|}{2}.
\end{equation}
Therefore (at any point in the spacetime)
\begin{equation}
\kappa_\mathrm{generator} = \frac{|Q_\chi|}{2} = \Big|(u\cdot \chi)(a\cdot s) -(s\cdot \chi)K_0\,\Big|.
\end{equation}
At the universal horizon, $u\cdot \chi=0$ by definition, and thus $\chi$ and $s$ are parallel. Therefore
\begin{equation}\label{kappauh}
\left.\kappa_\mathrm{generator}\right|_\mathrm{UH}=K_{0|\mathrm{UH}}\;  \Vert\chi\Vert_\mathrm{UH},
\end{equation}
where the absolute value and the explicit minus sign can safely be removed given that both $K_0$ and $(s\cdot\chi)$ are both positive on the universal horizon.
Indeed, this is consistent with  \cite{Berglund:2012bu} from which, by confronting our equation (\ref{killingderiv}) with equation (22) of \cite{Berglund:2012bu}, one can deduce $\kappa_\mathrm{generator}=Q_\chi/2$. 
In closing let us stress that this derivation relies \emph{very heavily} on the special symmetries of the solution and that $ ||\chi||_\mathrm{UH}\neq 0$ on the universal horizon.

%------------------------------------------------------------------------------------------------------------------------------------------
\subsubsection{Peeling}
%------------------------------------------------------------------------------------------------------------------------------------------

A specific class of spherically symmetric black holes was examined in \cite{Barausse:2011pu}, which in Eddington--Finkelstein coordinates take the form
\begin{equation}
\d s^2 = -e(r) \,\d \nu^2 +2f(r)\,\d\nu\,\d r +r^2\,\d\Omega.
\end{equation}
First, in analogy with section (\ref{peeling}), change this into Schwarzschild coordinates. 
Set
\begin{equation}
\d t = \d \nu -\frac{f(r)}{e(r)}\,\d r,
\end{equation}
so that
\begin{equation}
\d s^2 = -e(r)\, \d t^2 +\frac{f(r)^2}{e(r)}\,\d r^2 +r^2\d\Omega.
\end{equation}
Consider an out-going null ray
\begin{equation}
e(r) \d t^2 =  \frac{f(r)^2}{e(r)}\,\d r^2,
\end{equation}
so that
\begin{equation}
\frac{\d r}{\d t}= \frac{e(r)}{f(r)}.
\end{equation}
For $r_1$ and $r_2$ close to the universal horizon at $r=r_{UH}$ 
\begin{equation}
\left.\frac{\d (r_1-r_2)}{\d t}= \frac{\d}{\d r}\left(\frac{e(r)}{f(r)}\right)\right\vert_{\mathrm{UH}} (r_1-r_2)+ O\left([r_1-r_2]^2\right),
\end{equation}
and so for a generic universal horizon we can define
\begin{equation}
\kappa_\mathrm{peeling}= \left.\frac{1}{2}\frac{\d}{\d r}\left(\frac{e(r)}{f(r)}\right)\right\vert_{\mathrm{UH}},
\end{equation}
in general. 

Let us now apply this construction to the simplest explicit example we can find. 
Taking a look at section (4.2) in reference~\cite{Berglund:2012fk}, we see an example of an exact solution with
\begin{equation}
e \left( r \right) =1-{\frac {r_0}{r}}-{\frac {r_u \left( r_0+r_u \right) }{{r}^{2}}}; 
\qquad f(r)=1.
\end{equation}
Here
\begin{equation}
r_u=\left( \sqrt{C}-1\right) \frac{r_0}{2},
\end{equation}
with $C$ a constant depending on the coupling constants of the theory. 
Plugging this into the above, we find, that for this specific example
\begin{equation}
\kappa_\mathrm{peeling}=\frac{2C}{r_0}.
\end{equation}
Berglund \emph{et al.} \cite{Berglund:2012bu} compute the equivalent of
\begin{equation}
 \frac{Q_\chi}{2}=\frac{2C}{r_0},
\end{equation}
Thus, (at least in situations where they can both meaningfully be defined), $\kappa_\mathrm{peeling} = \kappa_\mathrm{generator}$ for universal horizons. 
We do not wish to apply this construction to the general solutions in terms of asymptotic expansions presented in that paper, as those are only valid for large $r$, and as such, ill-adapted to this calculation.
%

%------------------------------------------------------------------------------------------------------------------------------------------
\subsubsection{Null normal}
%------------------------------------------------------------------------------------------------------------------------------------------

Let us now see if it is possible to extend the notion $\kappa_\mathrm{normal}$ to a universal horizon, at least in a highly symmetric case. 
First, define a vector $\lambda$, everywhere orthogonal to $\chi$, by
\begin{equation}
\lambda^a=(s\cdot \chi)u^a -(u\cdot \chi)s^a.
\end{equation}
(There is a sign ambiguity in this definition depending on whether you want the inwards or outwards pointing unit spacelike vector at infinity.) 
Note also, that on the \emph{Killing} horizon, $u\cdot \chi=s\cdot \chi$, so $\lambda^a =\chi^a$.
Now, by spherical symmetry
\begin{eqnarray}
\nabla_a(\chi^2)&=& \nabla_a(\chi^b)\chi_b+\chi^b\nabla_a\chi_b \nonumber\\
&=& -\frac{Q_\chi}{2}\chi_b(u_as^b-s_au^b)-\frac{Q_\chi}{2}\chi^b(u_as_b-s_au_b) \nonumber \\
&=&Q_\chi\,(u\cdot\chi)s_a-Q_\chi\,(s\cdot\chi)u_a \nonumber\\
&=&- Q_\chi \lambda_a \nonumber\\
&=& -2\kappa_\mathrm{generator} \; \lambda_a,
\end{eqnarray}
everywhere in the spacetime. 
Off the Killing horizon, this seems to provide the most natural \emph{definition} of $\kappa_{\mathrm{normal}}$, and it is equal to $\kappa_\mathrm{generator}$.

%------------------------------------------------------------------------------------------------------------------------------------------
\subsubsection{Inaffinity}
%------------------------------------------------------------------------------------------------------------------------------------------

Likewise, for null horizons we have defined $\kappa_{\mathrm{inaffinity}}$ by
\begin{equation}
\chi^b\nabla_b\chi_c =\kappa_{\mathrm{inaffinity}} \chi_c.
\end{equation}
But, (as we have already seen), by spherical symmetry, 
\begin{equation}
\nabla_a\chi_b=-\frac{Q_\chi}{2} (u_a s_b-u_bs_a),
\end{equation}
so, now evaluating on the universal horizon, we have
\begin{eqnarray}
\chi^b\nabla_b\chi_c&=&- \frac{Q_\chi}{2} \chi^b(u_bs_c-u_cs_b) \nonumber\\
&=&-\frac{Q_\chi}{2}\{ (u\cdot \chi)s_c-(s\cdot \chi)u_c\} \nonumber \\
&=&\frac{Q_\chi}{2} \;\lambda_c \nonumber \\
&=& \kappa_{\mathrm{inaffinity}} \; \lambda_c,
\end{eqnarray}
where the last line is our \emph{definition} of $\kappa_\mathrm{inaffinity}$, which is now seen to be the same as $\kappa_\mathrm{normal}$ and $\kappa_\mathrm{generator}$.

%------------------------------------------------------------------------------------------------------------------------------------------
\subsubsection{Tension in a rope}
%------------------------------------------------------------------------------------------------------------------------------------------

Note that it is not at all obvious there should be any possible calculation for the tension in a rope at infinity, as our universal horizon is inside the Killing horizon, where nothing can stay still, so we certainly must abandon the notion of $\kappa_\mathrm{tension}$ directly relating to the tension on a rope held at infinity.

However, if we want to mathematically continue this idea, we want to calculate
\begin{equation}
\kappa_\mathrm{tension}^2 = \lim_\mathrm{UH} \left( \Vert \chi\Vert ^2 \; \Vert A\Vert ^2 \right)
\end{equation}
Because the universal horizon is not a null surface the limit is straightforward, and it is easy to see that
\begin{equation}
\kappa_\mathrm{tension}^2 = \lim_\mathrm{UH} \left\{\frac{-(\chi^b\nabla_b\chi^c)(\chi^a\nabla_a\chi_c)}{-\chi^a\chi_a} \right\}.
\end{equation}
But, we can again use equation (\ref{killingderiv}), so that
\begin{equation}
\kappa_\mathrm{tension}^2=\left.\frac{1}{4}
\frac{-Q_\chi^2\{(\chi \cdot u)s_a-(\chi \cdot s)u_a\}\,\{(\chi \cdot u)s^a-(\chi \cdot s)u^a\}}
{-\chi^2}\right|_{UH}=\left.\frac{Q_\chi^2}{4}\right|_\mathrm{UH}.
\end{equation}
That is
\begin{equation}
\kappa_\mathrm{tension} = \frac{|Q_\chi|}{2}. 
\end{equation}
Again we note that many of these definitions degenerate.

%------------------------------------------------------------------------------------------------------------------------------------------
\subsubsection{2-d expansion}
%------------------------------------------------------------------------------------------------------------------------------------------

Another possibility is to consider the quantity defined by Jacobson and Parentani~\cite{Jacobson:2008cx}. 
Instead of the form given in that paper, for spacelike regions (such as we have under consideration here) it is better to start from the basic definition 
\begin{equation}
\theta_{2d}=\frac{\frac{1}{2}u^a\nabla_a \chi^2}{\chi^2-(\chi\cdot u)^2},
\end{equation}
use the fact that $(\chi\cdot u)\to 0$ on the universal horizon, and expand the numerator to obtain
\begin{equation}
\left.\theta_{2d}\right|_\mathrm{UH} = \frac{u^a\chi^b\nabla_a\chi_b}{\chi^2}.
\end{equation}
Now we can again use our useful symmetries, and note that on the universal horizon $\chi^2 = (\chi\cdot s)^2$, to see
\begin{equation}
\left.\theta_{2d}\right|_\mathrm{UH} =  \frac{-Q_\chi u^a \chi^b (u_as_b-u_bs_a)}{2\,\chi^2}=\frac{Q_\chi\,(\chi\cdot s)}{2(\chi\cdot s)^2}
 = {Q_\chi\over 2(\chi\cdot s)}.
\end{equation}
We see that, whereas for Killing horizons, where we relate this quantity to the surface gravity through normalization with an appropriate horizon-crossing timelike vector $\chi\cdot u$, here we want to normalize with an appropriate spacelike vector $\chi\cdot s$.  
Specifically, for universal horizons we can define
\begin{equation}
\kappa_\mathrm{expansion} = \left.\left\{(\chi\cdot s) \,\theta_{2d}\right\}\right|_\mathrm{UH} =  \frac{Q_\chi}{2}.
\end{equation}
In particular, comparing with equation (\ref{kappauh}), we see that $\theta_{2d}= K_0$ at the universal horizon.

%------------------------------------------------------------------------------------------------------------------------------------------
\subsubsection{Summary}
%------------------------------------------------------------------------------------------------------------------------------------------

For a spherically symmetric universal horizon, and subject to the definitions adopted above, we have
\begin{equation}
\kappa_\mathrm{generator} = \kappa_\mathrm{normal} = \kappa_\mathrm{inaffinity}=\kappa_\mathrm{tension}=  \kappa_\mathrm{expansion}.
\end{equation}
When it is possible to calculate $\kappa_\mathrm{peeling}$, we find  $\kappa_\mathrm{peeling}=\kappa_\mathrm{generator}$.

Note that it is only by using tricks of spherical symmetry that we have been able to define some extension of $\kappa_\mathrm{normal}$ and $\kappa_\mathrm{inaffinity}$. 
The most natural notions for such horizons seem to be $\kappa_\mathrm{generator}$, $\kappa_\mathrm{expansion}$ and $\kappa_\mathrm{peeling}$, as they do not \emph{a priori} require a null surface. 
In the case of our modified gravity scenario, the symmetries of the problem seem to have reduced the plethora of surface gravities we have. 
Likewise, in analogue cases, if we have enough symmetry in the set up, the number of distinct surface gravities should collapse.

Indeed the calculations presented in this section rely so heavily on the spherical symmetry, that for a stationary non-static solution possessing a universal horizon, it seems that a completely new mode of attack would need to be developed. 
It is far from obvious which if any of these degeneracies would remain in such a case, and it seems somewhat  unlikely that the notions of $\kappa_\mathrm{inaffinity}$ and  $\kappa_\mathrm{normal}$ could be developed at all. 
Overall, the best statement seems to be this: There are many possible  definitions of surface gravity, identical in cases of high symmetry. 

%------------------------------------------------------------------------------------------------------------------------------------------
\section{Discussion}
%------------------------------------------------------------------------------------------------------------------------------------------

In this paper we have considered a number of different definitions of surface gravity, all of which reduce to the standard case in stationary general relativity. 
We have shown in the case of stationary analogue black holes how these different surface gravities can be calculated, and how they are related. 

These concerns are also important for modified gravity, and we have illustrated this with one example involving the so-called ``universal horizon''. 
In general,  the differences between these definitions, and appropriate choices of which to use, will become more relevant the less symmetry there is in the case under consideration. 
The symmetries in question might be obvious ones (spherical symmetry, axial symmetry), but might also be less obvious --- such as the enhanced conformal symmetry at general relativity horizons that is at least partly connected with the specific field equations and inter-twined with the rigidity theorem and zeroth law. 

Once one moves away from standard general relativity the situation becomes \emph{much} more complicated than one might have naively expected. 

\section{Acknowledgements}
We thank Thomas Sotiriou for helpful comments.

%------------------------------------------------------------------------------------------------------------------------------------------

%------------------------------------------------------------------------------------------------------------------------------------------
%------------------------------------------------------------------------------------------------------------------------------------------
\end{document}